\documentclass[preprint,aps]{revtex4}

\usepackage{graphicx}
\usepackage{dcolumn}
\usepackage{bm}

\begin{document}



\title{A finite-temperature liquid-quasicrystal transition in a lattice model}

\author{Z. Rotman}
\author{E. Eisenberg}

\affiliation{Raymond and Beverly Sackler School of Physics and Astronomy,
Tel Aviv University, Tel Aviv 69978, Israel}



\begin{abstract}
We consider a tiling model of the two-dimensional square-lattice, where each site is tiled with one of the sixteen Wang tiles. The ground states of this model are all quasi-periodic. The systems undergoes a disorder to quasi-periodicity phase transition at finite temperature. Introducing a proper order-parameter, we study the system at criticality, and extract the critical exponents  characterizing the transition. The exponents obtained are consistent with hyper-scaling.
\end{abstract}


\maketitle
It has been known for a while that the two- or three-dimensional space may be tiled by ordered but aperiodic tilings, in addition to periodic lattice structures \cite{Grunbaum-Shephard}. This aperiodic order is realized in nature by certain alloys, called quasicrystals
\cite{shechtman,Lifshitz03}, which are believed to exhibit in thermal equilibrium aperiodic crystalline order \cite{janot,Lifshitz07} -- long-range positional order lacking any periodicity. Aperiodic tiling models are extensively used to analyze quasycrystals at zero or finite temperature. However, understanding of the transition region between the disordered (fluid) phase and the quaicrystal phase is incomplete yet.

\begin{figure}
\includegraphics[width=9cm,height=6.9cm,angle=0]{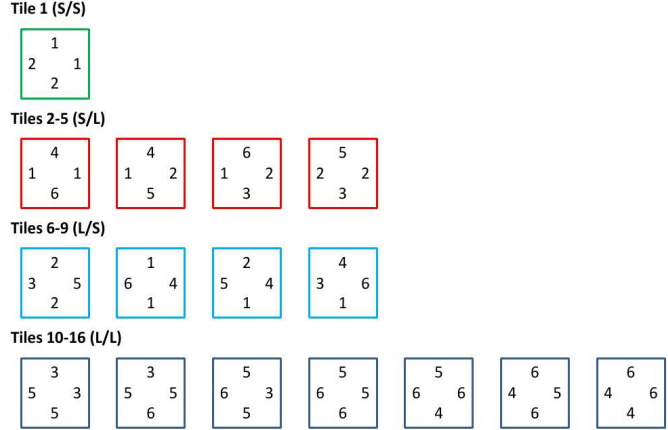}
\caption{(Color Online) The sixteen Amman tiles. The four rows correspond to the four types of tiles (see text).} \label{fig-tiles}
\end{figure}

Here we consider a model of interacting tiles on a square, two-dimensional, lattice. The model has been previously studied in \cite{lp,koch}. Each site of the square lattice is tiled with a tile, characterized by four labels attached to its edges. The labels take one of possible six labels (or colors). The interaction is with nearest neighbor tiles, and the bond energy is zero if the labels of both neighboring edges match, or one otherwise. It was found by Ammann\cite{book} that if one limits the allowed tiles to a group of $16$ tiles (out of all the possible $6^4$ tiles), then all zero energy states of the model (also known as perfect tilings) are non-periodic. The $16$ Amman tiles are presented in figure \ref{fig-tiles}. The perfect tilings are then the ground states of this system. Their non-periodicity can be shown using a mapping of the six tile labels into one of two symbols: $S$ for labels $\{1,2\}$ and $L$ for labels $\{3,4,5,6\}$.
Since all Amman tiles have labels belonging to the same set ($S$ or $L$) on
both horizontal (vertical) edges, the mapping thus classifies the $16$ tiles into $4$ types according to their set along each axis: $\{1\}$ S-S, $\{2,3,4,5\}$ S-L, $\{6,7,8,9\}$ L-S and $\{10,11,12,13,14,15,16\}$ L-L.
Identifying the symbols $S$ and $L$ with the short (S) and long (L) Fibonacci tiles, it follows from the properties of the Amman tiles that any perfect tiling is mapped into a two-dimensional square Fibonacci tiling \cite{ron}, thus aperiodic.

 The finite temperature behavior of this tiling model was studied numerically in  \cite{lp}. The model has multiple ground states (uncountable infinite number for the infinite plane), all are aperiodic, and thus its dynamics upon fast cooling was suggested to be a model for glassiness. The lattices studied in \cite{lp}were in the range $8\leq N\leq32$ ($N$ being the lattice linear size) with free boundary conditions. Numerical results supported the existence of a phase transition, measured by a growing peak in the specific heat, and the transition was concluded to be of second order. Recently \cite{koch}, it was shown that the phase transition observed in \cite{lp} is a disorder (fluid) to quasicrystal transition. Phase transition analysis in \cite{koch} followed an analytical approach supported by numerical simulations. The transition was studied using the overlap of a configuration with a ground state $\gamma$. The fraction of tiles in a configuration $c$ matching a ground state $\gamma$ is denoted $\phi(c,\gamma)$. The overlap, normalized by its averaged over all ground states $\psi(c)=\int\phi(c,\gamma)d\lambda(\gamma)$, is then thermally averaged to yield $Q_\beta$
\begin{equation}
\label{overlap} Q_\beta(\gamma)=\frac{1}{Z}\int{\frac{\phi(c,\gamma)e^{-\beta H(c)}}{\psi(c)}},
\end{equation}
where the integration is over configurations $c$ (ensemble averaging), $H(c)$ is the energy of the configuration, and $Z$ is the partition function $Z=\int \exp(-\beta H(c))$.
The high temperature limit of $Q_\beta$ is, by normalization, $1$.

To account for the infinite number of ground states, Koch and Radin defined the quantity $q_{RK}$
\begin{equation}
\label{kochop} q_{RK}(\beta)=-\int Q_\beta(\gamma) \ln[Q_\beta(\gamma)] d\lambda(\gamma),
\end{equation} and then $q_{RK}(\beta << \beta_c)=0$.
Analytical calculation showed that $q_{RK}$ vanishes identically for sufficiently high finite temperatures. Numerical simulations were then used to show $q_{RK}$ does not vanish for low temperatures, thus proving the existence of a transition. Simulations persented in \cite{koch} employed fixed boundary conditions corresponding to a specific ground state $\sigma$. Then, the overlap of the configuration with the chosen ground state $\sigma$ was used to approximate $q_{RK}$ at low temperatures, where it is expected to contribute dominantly. Looking at system sizes $32\leq N\leq 256$, they concluded that the transition is of third or higher order. In addition, it was suggested that the transition has no renormalization fixed point.

Square Fibonacci quasicrystals are known to exhibit distinct delta-function peaks in the fourier transform, in a similar fashion to the peaks observed in crystalline solid \cite{ron}. We therefore suggest to use the weights of these distinct peaks to define an order parameter for the disorder-to-quasicrystal transition, one that is simpler and easier to access numerically then $q_{RK}$. Many frequencies show a peak, all are irrational and related to the 'golden mean' $\tau=(1+\sqrt{5})/2$. Frequency chosen for this study is $\vec{k_0}=\bigl((\tau-1)/\tau,(\tau-1)/\tau\bigr)$, but results are similar for all related frequencies. The (complex) order parameter is then the amplitude of the peak at $\vec{k_0}$
\begin{equation}
\label{op} q= \int e^{i\vec{k_0}\cdot \vec{r}}\delta_{\sigma_{\vec{r}},1}\vec{dr}.
\end{equation} Tile $1$ is chosen for simplicity, as it is the only tile which is the only one of its type, i.e. it is the only tile with both edges being S edges. Similar results are obtained using any other type e.g tiles 2-5, 6-9 or 10-16.

In the following we present numerical simulations for $10\leq N\leq 400$ and free boundary conditions. Simulations starts from a ground state and then thermalized at the desired temperature. Thermalization during cooling was found to be significantly less efficient. As expected, the order-parameter auto-correlation times measured near criticality are extremely long, and thus we took each run to last $10^8$ Monte-Carlo Steps.  To account for multiple ground states and to assure good cover of phase space we repeated the simulations starting from different ground states. Finite size scaling analysis was used to calculate the critical temperature and critical exponents of the phase transition.

\begin{figure}
\includegraphics[width=6.9cm,height=6.9cm,angle=-90]{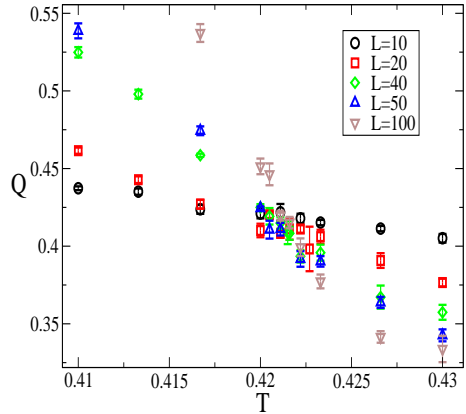}
\caption{(Color Online) Binder cumulant of the tiling model. The collapse of crossings for various lattice sizes signatures the location of the critical point.} \label{fig-bin}
\end{figure}

\begin{figure}
\includegraphics[width=6.9cm,height=6.9cm,angle=-90]{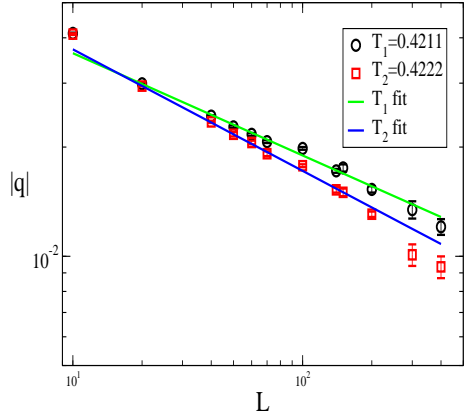}
\caption{(Color Online) Order parameter measurements performed at $T_1$ and $T_2$. Power law behavior is fitted, L=10 is excluded from the analysis. Fitted exponents are $0.28$ and $0.33$.} \label{fig-q}
\end{figure}

\begin{figure}
\includegraphics[width=6.9cm,height=6.9cm,angle=-90]{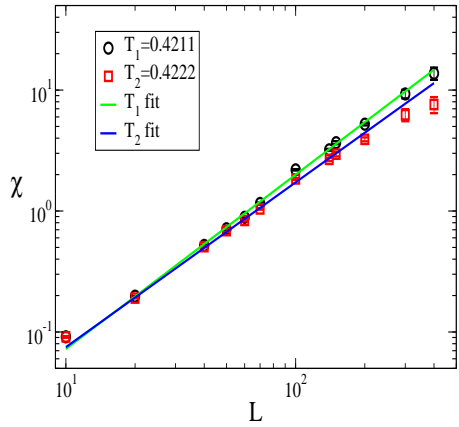}
\caption{(Color Online) Susceptibility $\chi$ fitted to a power law behavior at $T_1$ and $T_2$. Fitted exponents are $1.44$ and $1.36$.} \label{fig-chi}
\end{figure}

In order to determine the critical temperature, we analyzed the Binder cumulant $$Q=1-\frac{\langle |q|^4\rangle }{\langle 3|q|^2\rangle }$$ where $\langle \cdots \rangle$ denotes ensemble average. Crossings of $Q$ for system sizes $10\leq N\leq 100$, presented in figure \ref{fig-bin}, show no significant finite size effects within our accuracy. The relatively large error estimates due to the long relaxation processes allow for a moderate accuracy in fixing the critical temperature, which is estimated to be $T_c = 0.4216(5)$ (here and below the number in paranthesis is the uncertainty in the last digit). The critical temperature found is in agreement with previous estimates: $T_c=0.42(1)$ \cite{lp} and $\beta_c=1/T_c\approx2.4$ \cite{koch}.
We then studied larger lattices at two temperatures near criticality, $T_1=0.4211$ and $T_2=0.4222$. Finite size scaling of thermodynamic quantities at $T_c$ provides estimates for the critical exponents. In the following we verify that the critical exponents obtained for $T_1$ and $T_2$ are similar, in order to ensure that our inaccuracy in $T_c$ does not take us out of the critical regime for the lattice sizes studied. The order parameter $|q|$ is expected to scale as $\langle|q|\rangle (L) \sim L^{-\beta/\nu}$.
Figure \ref{fig-q} presents numerical results from measurements at $T_1$ and $T_2$ and fits to the power-law form.
Based on the two fits we estimate $\beta/\nu=0.30(3)$. Similarly, the susceptibility defined by
\begin{equation}
\label{chi}\chi=\frac{N^2(\langle |q|^2\rangle -\langle |q|\rangle ^2)}{T},
\end{equation}
was also measured at $T_1$ and $T_2$ and fitted to the scaling form $\chi\sim L^{\gamma/\nu}$ (figure \ref{fig-chi}). Estimation of $\gamma/\nu$ from the two fits leads to $\gamma/\nu=1.40(5)$. Note that these two independent measurements satisfy the hyperscaling relation $2\beta/\nu+\gamma/\nu=d$.

\begin{figure}
\includegraphics[width=6.9cm,height=6.9cm,angle=-90]{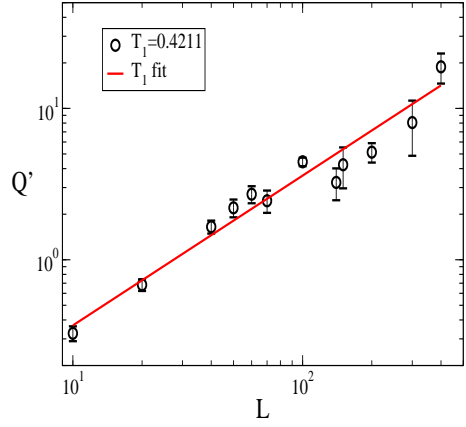}
\caption{(Color Online) Measurements of $Q'$ for $T_1=0.4211$. Fitted exponent is $0.99$.} \label{fig-dbin}
\end{figure}

\begin{figure}
\includegraphics[width=6.9cm,height=6.9cm,angle=-90]{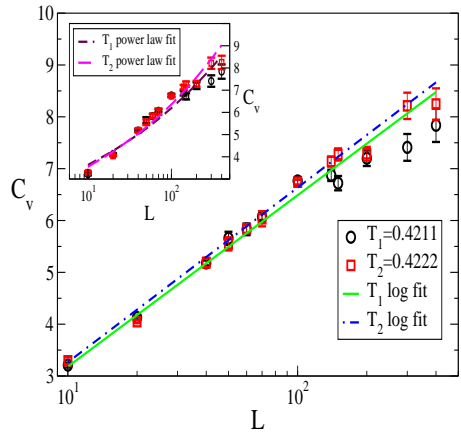}
\caption{(Color Online) Specific heat measurements at $T_1$ and $T_2$. Data is well fitted by a logarithmic function. For comparison, best power law fits are presented in the inset, yielding exponents $0.23$ and $0.25$.} \label{fig-cv}
\end{figure}

In order to estimate the critical exponent $\nu$, we studied the derivative of the binder cumulant ($Q$) with respect to  the inverse temperature $\beta$. Based on finite-size-scaling arguments $$Q'=\frac{\partial Q}{\partial \beta}\bigg|_{T=T_c}$$ is expected to diverge like $L^{1/\nu}$. It is easy to see that this derivative is obtained from the energy and order-parameter moments in the following form
$$Q'=\frac{2\langle q^4\rangle \langle q^2E\rangle /\langle q^2\rangle -\langle q^4E\rangle -\langle q^4\rangle \langle E\rangle }{3\langle q^2\rangle ^2}. $$
Measurements presented in figure \ref{fig-dbin} indeed show a power law behavior of $Q'(N)$, and the  exponent estimated is $1/\nu=1.05(15)$. This value for $\nu$ is consistent with measurements of the specific heat
\begin{equation}
\label{cv}C_v=\frac{\langle E^2\rangle -\langle E\rangle ^2}{N^2 T^2},
\end{equation}
which are best-fitted by a logarithmic growth (figure \ref{fig-cv}),
i.e. $\alpha=2-d\nu=0$, or $\nu=1$. The results are consistent with the
analysis performed in \cite{lp}, leading to $\nu=1.6(5)$.

In conclusion, we show that the use of the Bragg peak allows for an analysis of the disorder to quasicrystal transition in a two-dimensional lattice model based on Amman tiles. The transition occurs at a finite temperature and is of a second order. Critical exponents were measured and shown to satisfy hyperscaling relations. This model is therefore suitable for study of the critical emergence of quasi-periodic order.

\begin{acknowledgements}
We are grateful to Ron Lifshitz for important discussions and insightful comments and to Hans Koch for providing useful information regarding the numerical simulations as well as a critical reading of the manuscript.
\end{acknowledgements}

\end{document}